\documentclass{article}
\usepackage{spconf,amsmath,graphicx,amsthm,hyperref,bbm,pgfplots,tikz, booktabs}
\usepackage[shortlabels]{enumitem}
\usepackage{subcaption}
\usetikzlibrary{pgfplots.groupplots}

\title{Hierarchical Prosody Modeling \\ for Non-Autoregressive Speech Synthesis}
\name{Chung-Ming Chien \quad Hung-yi Lee}
\address{
College of Electrical Engineering and Computer Science, National Taiwan University, Taiwan}
\begin{document}
%
\maketitle

\begin{abstract}
Prosody modeling is an essential component in modern text-to-speech (TTS) frameworks. By explicitly providing prosody features to the TTS model, the style of synthesized utterances can thus be controlled. However, predicting natural and reasonable prosody at inference time is challenging. In this work, we analyzed the behavior of non-autoregressive TTS models under different prosody-modeling settings and proposed a hierarchical architecture, in which the prediction of phoneme-level prosody features are conditioned on the word-level prosody features. The proposed method outperforms other competitors in terms of audio quality and prosody naturalness on objective and subjective evaluation.
\end{abstract}


\vspace{-1.7mm}
\section{Introduction}
\label{sec:intro}
\vspace{-1.2mm}

In order to synthesize human-like speech utterances by Text-to-speech (TTS), it is important to model the variation in speech signals, including rhythm, intonation, and stress, etc.
These factors, collectively referred to as \textit{prosody}, are not contained in the text transcripts, but are very important for conveying information that is not specified by the texts.
Providing additional prosody information to the TTS model is referred to as \textit{prosody modeling}, which enables expressive and controllable speech synthesis \cite{Skerry-Ryan2018, Wang2018, Zhang2019, Hsu2019, Lee2019}. 

In this work, we focus on fine-grained prosody modeling \cite{Lee2019}, where the prosody of an utterance is represented as a sequence of prosody features instead of a single sentence-level prosody feature.
Each fine-grained prosody feature encodes the prosody associated with a speech segment, such as a phoneme or a word. 
Aside from enabling local prosody control in speech synthesis, fine-grained prosody modeling also further reduces the complexity of the TTS task itself, since the information contained in each fine-grained prosody feature is explicitly assigned to a speech segment.
This makes prosody modeling especially an important component in non-autoregressive TTS framework \cite{Ren2020, Lancucki2020}.
Compared with autoregressive models that suffer from the false alignment problem, non-autoregressive TTS systems are faster and more robust \cite{Ren2019}, but the training is much more difficult since the model has to predict the entire mel-spectrogram simultaneously.
The prosody features provide additional information to the TTS model and effectively simplify the mapping between the text input and the speech output.

Many handcrafted features, such as fundamental frequency (F0) contour and energy \cite{Ren2020, Lancucki2020}, and even neural-based features \cite{Skerry-Ryan2018, Wang2018, Zhang2019, Hsu2019, Lee2019, Sun2020g}, can be used to model the prosody variation within an utterance.
The prosody features can be extracted from ground-truth speech signals at training time, while \textit{how to generate natural prosody features at inference time} remains an open problem.
Some proposed to infer prosody features from phoneme-level features \cite{Ren2020, Lancucki2020, Sun2020g, Stanton2018}.
However, we consider that the attributes that affect the prosody of an utterance, such as the meaning of the sentence, and the speaker's intention or sentiment, can be better realized with word-level features rather than the phoneme-level ones. 
The effect of different granularity in fine-grained prosody modeling is studied in this work, and the experimental results verify the above hypothesis.

There are three major contributions of this paper. 
First, we figure out that there is a trade-off between the quality of synthesized audio samples and the accuracy of prosody prediction, with respect to the granularity of fine-grained prosody modeling.
Second, we compare different approaches to extracting prosody features in terms of audio quality and prosody naturalness.
Last, we propose a hierarchical prosody modeling framework, where phoneme-level prosody prediction is conditioned on word-level prosody prediction, to combine the advantage of phoneme-level and word-level prosody modeling.
With both objective and subjective evaluation, we verify that the proposed hierarchical model outperforms any other prosody modeling paradigms of interest.
The readers are encouraged to listen to the audio samples attached here
\footnote{Audio sample: \href{https://ming024.github.io/hierarchical_prosody_modeling/}{ \texttt{https://ming024.github.io/hierarchical\\\_prosody\_modeling}}}.

\vspace{-1.7mm}
\section{Background}
\label{sec:background}
\vspace{-1.2mm}

\begin{figure}[tb]
    \centering
    \centerline{\includegraphics[width=8.8cm]{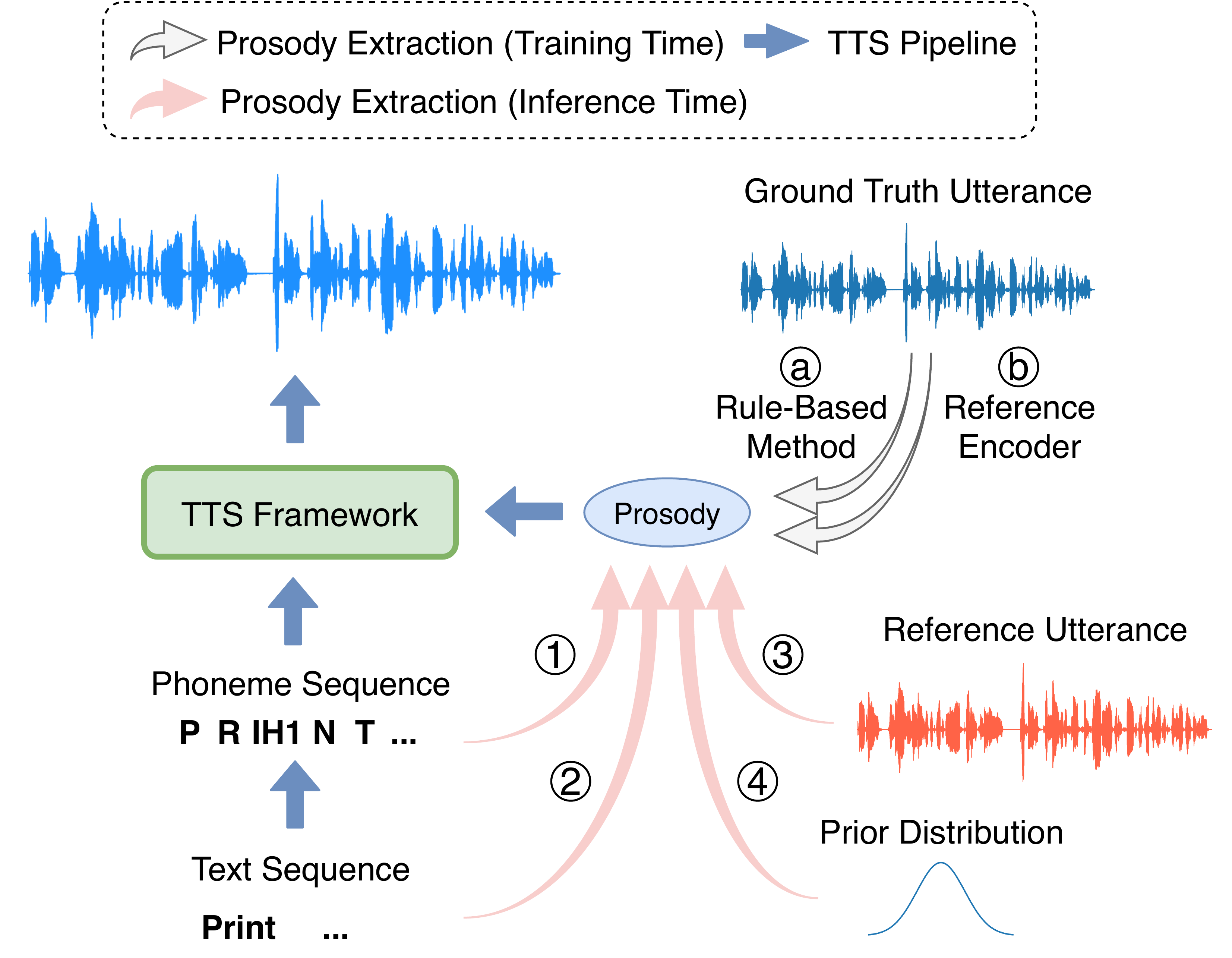}}
    \vspace{-1.7mm}
    \caption{General pipeline for TTS systems with prosody modeling.}
    \vspace{-3.2mm}
    \label{fig:TTS_with_prosody}
\end{figure}

A typical TTS model takes a phoneme sequence as inputs to predict mel-spectrogram or raw waveform.
As shown in Fig.~\ref{fig:TTS_with_prosody}, prosody features can serve as extra inputs of the TTS model, which simplifies the one-to-many mapping from the phoneme sequence to the speech signals.
At training time, the prosody features are extracted from the ground-truth, enabling the TTS model to generate speech signals with prosody similar to the ground-truth utterance.
The prosody features can be obtained with the following methods:
\begin{enumerate}[label=\textcircled{\footnotesize
\alph*}]
    \vspace{-0.7mm}
    \item Rule-based prosody features: F0 and energy can be computed with many off-the-shelf packages, and are often used to model prosody variation \cite{Ren2020, Lancucki2020}.
    \vspace{-0.7mm}
    \item Neural-based prosody features: a reference encoder is used to drain information from the ground-truth utterance \cite{Skerry-Ryan2018, Wang2018, Zhang2019, Hsu2019, Lee2019, Sun2020g}. 
    Suppose the size of the representation is small enough. In that case, it is be expected that the information extracted from the ground-truth is mostly about prosody and other information that is not contained in the phoneme features, since the TTS model can obtain phonetic information from the input phoneme sequence \cite{Skerry-Ryan2018}.
\end{enumerate}

\vspace{-0.7mm}
At inference time, because we do not know the prosody of the synthesized audio in advance, the prosody features are generated by approaches different from in the training phase. 
There are many methods to generate the prosody features. 
We summarize them into four categories, as shown in Fig. \ref{fig:TTS_with_prosody}:
\begin{enumerate}[label=\textcircled{\footnotesize\arabic*}]
    \vspace{-0.7mm}
    \item Predicting prosody from the phoneme feature sequence \cite{Ren2020, Lancucki2020, Sun2020g, Stanton2018}.
    \vspace{-0.7mm}
    \item Predicting prosody from the word-level feature sequence. 
    There has been research trying to predict prosody features from the word-level features \cite{Talman2019}, whereas few works used prosody features predicted from word-level features to help speech synthesis.
    \vspace{-0.7mm}
    \item Imitating the prosody of a reference utterance, by conditioning the speech synthesis of arbitrary texts on prosody features extracted from the reference, usually used in prosody transfer tasks \cite{Skerry-Ryan2018, Wang2018, Zhang2019, Hsu2019, Lee2019}.
    \vspace{-0.7mm}
    \item Sampling prosody from a prior distribution \cite{Wang2018, Zhang2019, Hsu2019} 
\end{enumerate}
\vspace{-0.7mm}
This paper will focus on inferring prosody from text inputs, that is,  the phoneme and the word feature sequences.

\begin{figure*}[tb]
    \centering
    \centerline{\includegraphics[width=0.95\textwidth]{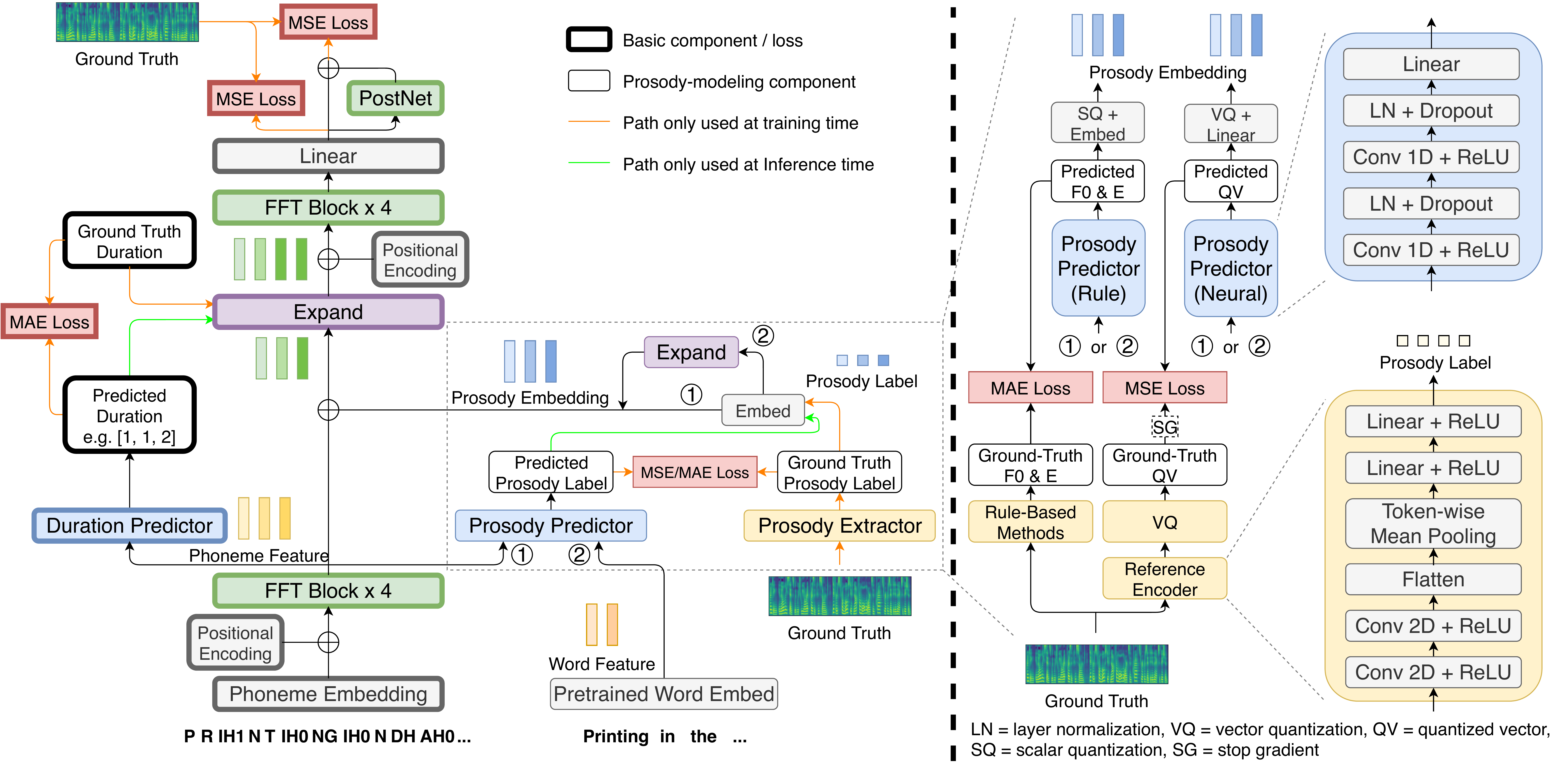}}
    \vspace{-1.7mm}
    \caption{\textbf{Left}: An overview of our model architecture. \textbf{Right}: A close-up view of the prosody-modeling components.}
    \vspace{-3.5mm}
    \label{fig:model_architecture}
\end{figure*}

It has been shown that hierarchical structures of prosody intrinsically exist in spoken language~\cite{Suni2017}.
Some previous work utilizes the hierarchical property of language to help speech synthesis.
\cite{Kenter2019, Wan2020} used an autoregressive model to predict frame-level prosody values and mel-spectrogram from hierarchical linguistic features.
However, their model only handles prosody prediction at frame-level, and the underlying prosody embedding is not fine-grained.
\cite{Sun2020h} proposed a TTS framework with hierarchical fine-grained prosody modeling, where the word-level and phoneme-level prosody features are encoded in two separate embedding spaces.
The previous work presented a multi-level prosody modeling structure, but the generation of prosody embedding at inference time is independent of the text inputs.
In our work, both word-level and phoneme-level prosody is modeled.
At inference time, the multi-level prosody is predicted from the phoneme-level and word-level features, which are derived from the input texts.
Our model is compatible with any pretrained word embedding, so the word-level prosody prediction benefits from the widely developed word representation models pretrained from large amounts of unlabeled texts.


\vspace{-1.7mm}
\section{Model Architecture}
\label{sec:model_architecture}
\vspace{-1.2mm}
In this section we introduce our model architecture based on FastSpeech 2 \cite{Ren2020}. 
The architecture comprises the basic components that are essential for generating valid speech signals and the prosody-modeling components, as shown in  Fig.~\ref{fig:model_architecture}.
The proposed hierarchical architecture is in Section \ref{ssec:hierarchical_prosody_modeling}.

\vspace{-1.7mm}
\subsection{Basic components}
\label{ssec:basic_components}
\vspace{-1.2mm}
The basic components in FastSpeech 2 include two Feed-Forward Transformer (FFT) \cite{Vaswani2017} stacks and a duration predictor. 
The configuration and hyper-parameters of the FFT blocks and the duration predictor in our implementation follow FastSpeech 2. 
The first FFT stack converts the phoneme sequence into a phoneme-level feature sequence in the resolution of the input sequence length. 
The duration predictor takes the phoneme-level feature sequence as inputs to predict the duration of each phoneme. 
Then the phoneme-level features are expanded in time according to each phoneme's duration to match the dimensionality of the output mel-spectrogram. 
At training time, ground-truth duration obtained with the Montreal Forced Aligner (MFA) \cite{McAuliffe2017} is used for expansion while at inference time predicted duration is used. 
The expanded feature sequence is then fed into the second FFT block, followed by a linear layer to predict mel-spectrogram. 

There are some differences between our implementation and the original FastSpeech 2.
We use an additional PostNet, which is the same as that used in Tacotron 2 \cite{Shen2018}, to post-process the output.
Mean Squared Error (MSE) loss is computed for both before- and after- post-processing output, while the original FastSpeech 2 uses Mean Absolute Error (MAE).
We also replace the MSE loss for duration prediction with MAE loss.
These minor modifications are beneficial to training efficiency and stability.
The entire network is trained in an end-to-end manner with the duration prediction loss and the mel-spectrogram prediction losses, including before- and after- post-processing parts.

\vspace{-1.7mm}
\subsection{Prosody extraction}
\label{ssec:prosody_extraction}
\vspace{-1.2mm}
The prosody-modeling components include the prosody extractors and the prosody predictors, as depicted in Fig.~\ref{fig:model_architecture}.
At training time, fine-grained prosody labels are extracted from the ground-truth mel-spectrogram.
Each prosody label contains the prosody information of a short speech segment (i.e., a word or a phoneme, collectively referred to as the \textit{token}).
The prosody labels are scalars or low-dimensional vectors.
They are mapped to a prosody embedding whose dimension equal to the dimension of the phoneme-level feature.
The prosody embedding sequence is then added to the phoneme-level feature sequence and fed into the second FFT stack for mel-spectrogram prediction.

We use two different configurations for prosody extractor, the rule-based one and the neural-based one, to extract the fine-grained prosody information.
We will go through their model architecture in detail in the following paragraphs.

\vspace{-1.7mm}
\subsubsection{Rule-based prosody extractors}
\label{sssec:rule_based_prosody_extractors}
\vspace{-1.2mm}
Following FastSpeech 2, we use F0 and energy as the rule-based prosody labels.
We use the DIO algorithm \cite{Morise2009} for F0 estimation and the L2-norm of each Short-Time Fourier Transform (STFT) frame for energy estimation to extract frame-level prosody labels.
The F0 and energy values are then averaged over the duration of each token to get token-level prosody labels, as proposed in \cite{Lancucki2020}.
The averaged values are then quantized into 256 bins, and transformed into a prosody embedding sequence by an embedding lookup\footnote{The only difference between the workflow here and FastSpeech 2 is that we use token-level instead of frame-level prosody labels.
However, in FastSpeech 2, the prosody labels are actually predicted from an expanded phoneme-level feature sequence, which only differs from the inputs of our token-level prosody prediction network by an expansion.}.


\vspace{-1.7mm}
\subsubsection{Neural-based prosody extractors}
\label{sssec:neural_based_prosody_extractors}
\vspace{-1.2mm}
Inspired by \cite{Sun2020g}, we use a Vector-Quantized Variational Autoencoder (VQ-VAE) ~\cite{van_den_Oord17} as the reference encoder.
The reference encoder is jointly learned with the TTS model to extract the prosody information from the ground-truth mel-spectrogram.
The reference encoder extracts a 3-dimensional latent representation for each token from the ground-truth mel-spectrogram.
The quantization in VQ-VAE serves as an information bottleneck, restricting the amount of information flow from the ground-truth to the mel-spectrogram prediction and thus encouraging the reference encoder to extract prosody information, which is not contained in the text inputs. 

The structure of the reference encoder can also be found in Fig. \ref{fig:model_architecture}.
The reference encoder comprises a stack of two 2D convolution layers, each composed of 32 filters with 3 $\times$ 3 kernel size and 1 $\times$ 1 stride.
A flatten layer followed, and a token-wise mean pooling is used to transform the frame-level feature sequence into a token-level feature sequence.
Then the following two linear layers projected into a 3-dimensional latent space.
All the layers are followed by a dropout layer with a 0.2 dropout rate.

A VQ codebook, consists of 256 codewords, is used to quantize the 3-dimensional latent vector to the nearest codeword (measured with L2 distance).
These prosody labels are passed to a linear layer to get prosody embeddings, which are then added to the phoneme feature sequence to predict mel-spectrogram.
A VQ loss is used to push the latent vectors and the codewords towards each other.

The quantized vectors from the reference encoder are treated as ground-truth prosody labels.
The prosody predictor is then learned to predict the ground-truth prosody labels from the token-level features derived from the text input.
The goal of the reference encoder is to provide the ground-truth prosody labels, so its parameters are fixed when training the prosody predictor, and are not used at inference time.

\vspace{-1.7mm}
\subsection{Prosody prediction}
\label{ssec:prosody_prediction}
\vspace{-1.2mm}
At inference time, since ground-truth prosody labels are not available, we have to predict the prosody of an utterance from the text input.
As described in Section \ref{sec:background}, both phoneme-level features and word-level features can serve as inputs for prosody prediction.
For word-level features, there are a variety of pretrained language models \cite{Joulin2016, Wolf2019} that can be used as a good word-embedding.
For phoneme-level features, we follow the configuration of FastSpeech 2, in which the output of the first FFT stack is used to predict the prosody labels.

The model architectures of the prosody predictors are shown in Fig. \ref{fig:model_architecture}, whose network architectures are the same as the duration model in FastSpeech 2.
When the prosody predictor takes word-level features as input, the predicted word-level prosody embedding sequence is expanded in time according to the word's phoneme number.
The prosody predictors are trained with MAE loss if rule-based prosody labels are used, and MSE loss if neural-based prosody labels are used, regardless of the input features.

\vspace{-1.7mm}
\subsection{Hierarchical prosody modeling}
\label{ssec:hierarchical_prosody_modeling}
\begin{figure}[tb]
\centering
\centerline{\includegraphics[width=8cm]{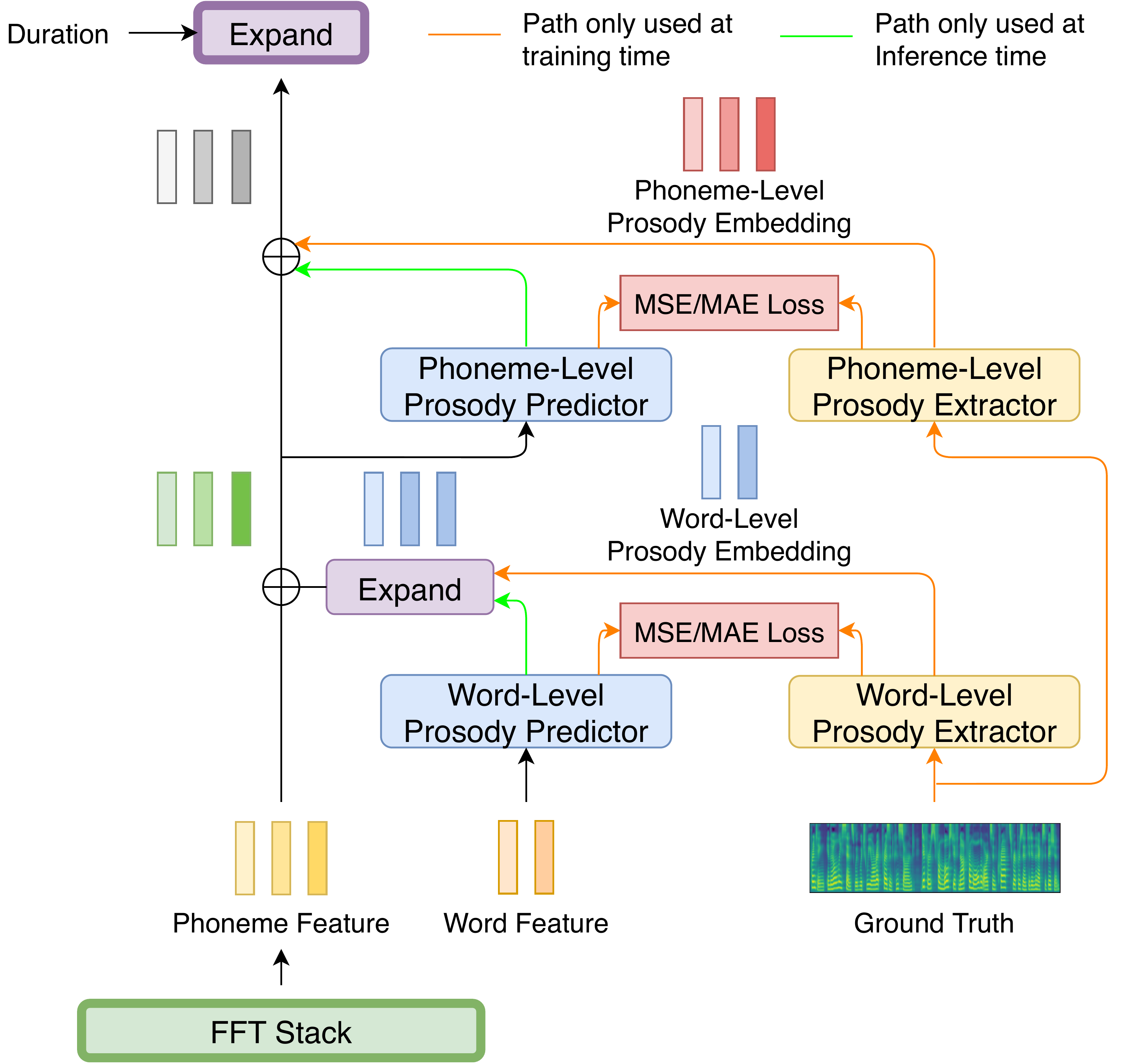}}
\vspace{-1.7mm}
\caption{Proposed hierarchical prosody modeling architecture.}
\vspace{-3.7mm}
\label{fig:hierarchical_prosody_predictor}
\end{figure}

\vspace{-1.2mm}
Under our model architecture, it is possible to select different combinations of inputs (phoneme features or word features) and targets (rule-based or neural-based prosody labels) to train the prosody predictor.
Our preliminary experiments showed that the word features provide more accurate prosody prediction than the phoneme features.
However, the low-resolution of the word-level prosody embeddings hinders the quality of synthesized audio samples, since the TTS model tends to predict a blurred mel-spectrogram if there is not enough information given.
As a result, we design a hierarchical prosody modeling architecture, which is composed of a concatenation of a word-level prosody predictor and a phoneme-level prosody predictor, as shown in Fig. \ref{fig:hierarchical_prosody_predictor}.

In the proposed architecture, word-level prosody is first predicted by the word-level prosody predictor and expanded to match the phoneme sequence length.
The expanded embedding sequence is then added to the phoneme feature sequence.
By feeding the summed feature sequence into the phoneme-level prosody predictor, the prediction of the fine-grained phoneme-level prosody is thus conditioned on the result of the coarse-grained word-level prosody prediction.

The proposed hierarchical prosody modeling benefits from high-resolution phoneme-level prosody labels and accurate word-level prosody prediction simultaneously at inference time.
In this framework, both rule-based and neural-based prosody labels can be used to model phoneme-level or word-level prosody.
We will compare different hierarchical paradigms with the non-hierarchical ones in Section~\ref{sec:experiments}. 

\vspace{-1.7mm}
\section{Experiments}
\label{sec:experiments}
\vspace{-1.2mm}

\vspace{-1.7mm}
\subsection{Setup}
\label{ssec:setup}
\vspace{-1.2mm}
All of our models are trained on the LJSpeech dataset \cite{Ito2017}, which contains 13100 English utterances spoken by a female speaker.
We keep 892 sentences (with document title LJ001, LJ002, and LJ003) for testing, and the remaining are used for training.
MFA is used to convert the transcripts into phoneme sequences and find the alignments between the phoneme sequences and the utterances.
The audio samples are converted to 80-dimensional mel-spectrograms for training, and the predicted mel-spectrograms are converted to raw waveform with a pretrained MelGAN vocoder \cite{Kumar2019}.
Because the source code of FastSpeech 2 has not been released when this paper was written, we used our implementation in the experiments \footnote{FastSpeech 2: \href{https://github.com/ming024/FastSpeech2}{\texttt{https://github.com/ming024/FastSpeech2}}}. 

At training time, the models are trained with the Adam optimizer \cite{Kingma2014} (with $\beta_1 = 0.9$, $\beta_2 = 0.98$, and $\epsilon=10^{-9}$) for 300k steps with batch size 16, and the learning rate scheduling proposed in \cite{Vaswani2017} is applied.
The only exception is the word-level prosody predictor.
Since the loss and gradient flow of the word-level prosody predictor are independent of all other modules, we train this module separately with the Adam optimizer (with learning rate $10^{-4}$, $\beta_1 = 0.9$, $\beta_2 = 0.98$, and $\epsilon=10^{-9}$) for 30k steps with batch size 16.


\vspace{-2.7mm}
\subsection{Predictability of prosody labels from different features}
\label{ssec:predictability_prosody_labels_from_different_features}
\vspace{-1.2mm}
This experiment is designed to compare the performance of predicting prosody labels from phoneme-level features and word-level features.
We use fastText \cite{Joulin2016} and pretrained BERT \cite{Wolf2019, Devlin2019} to generate word-level feature sequences, which are used as the input of the word-level prosody predictor.
For a fair comparison, the pretrained BERT model is treated as a static feature extractor without fine-tuning.
To further study the influence of context information in prosody prediction, we also use BERT's hidden features from different layers as the input of the word-level prosody predictor.

\begin{figure}[tb]
\centering
\centerline{\includegraphics[width=8.5cm]{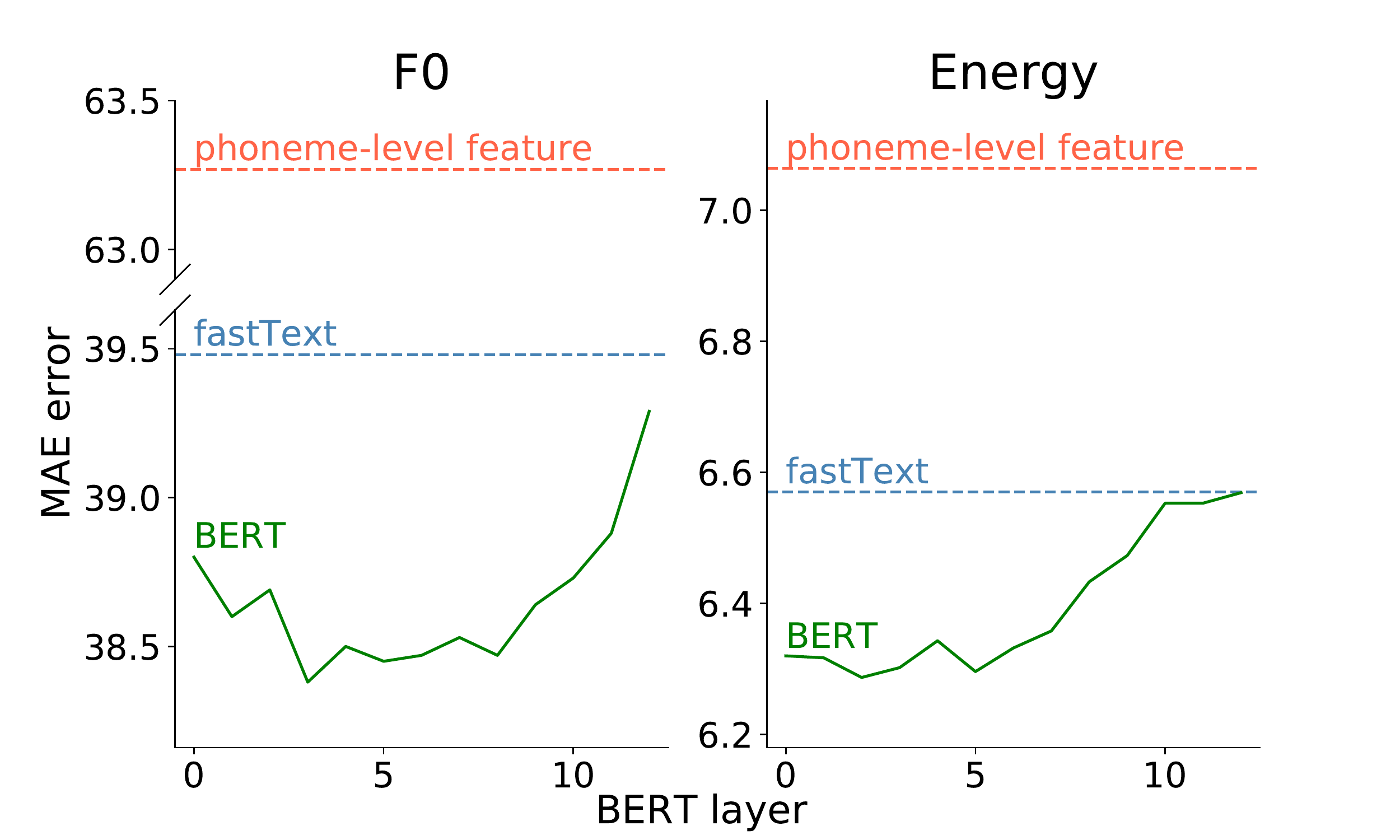}}
\vspace{-1.7mm}
\caption{MAE error of the F0 and energy prediction over the testing set, with phoneme-level features, fastText features and the outputs of the $0$-th to $12$-th layer of BERT as inputs.}
\vspace{-3.2mm}
\label{fig:prosody_prediction}
\end{figure}

The MAE loss of the predicted rule-based prosody labels in the testing set is reported in Fig. \ref{fig:prosody_prediction}.
All word-level features outperform the phoneme-level features, verifying the hypothesis that the word features can model prosody variation better than the phoneme features.
For different word embeddings, despite that BERT's performance is better than fastText, we cannot conclude that contextualized information helps prosody prediction.
Instead, we think contextualized information does not make any noticeable difference since the hidden features from high layers do not beat the low-layer features\footnote{The higher layers of BERT are generally considered to be more contextualized \cite{Ethayarajh2019}.}.
We will use the 0-th layer BERT features (i.e., the input word embedding of BERT) as the word-level features in all following experiments.
Similar results are also observed in neural-based prosody label prediction, which is not presented due to space limitation.

\begin{table}[tb]
    \setlength{\tabcolsep}{1.5mm}
    \begin{tabular}{cccccc}
        \toprule
        & GPE & VDE & FFE & F-MAE & E-MAE \\ 
        \toprule
        vanilla & .4063 & .2856 & .4493 & 42.829 & 8.205  \\
        \toprule
        P+R & .4084 & .2836 & .4660 & 41.806 & 7.363 \\
        \hline
        P+N & .4113 & .2898 & .4549 & 43.385 & 7.441 \\
        \hline
        P+N, rand. & .5278 & .3436 & .5278 & 57.119 & 9.290 \\
        \toprule
        W+R & .3952 & .2800 & .4498 & 40.202 & 7.264  \\
        \hline
        W+N & .3977 & .2972 & .4494 & 42.096 & 8.050 \\
        \hline
        W+N, rand. & .4759 & .2983 & .4824 & 48.924 & 8.265 \\
        \toprule
        H(W+R, P+R) & .3998 & .2812 & .4614 & 40.190 & 7.308  \\
        \hline
        H(W+R, P+N) & \textbf{.3886} & \textbf{.2758} & .4499 & \textbf{39.597} & 7.263 \\
        \hline
        H(W+N, P+R) & .3971 & .2832 & .4529 & 40.240 & \textbf{7.145}  \\
        \hline
        H(W+N, P+N) & .3994 & .2908 & \textbf{.4434} & 42.074 & 7.512  \\
        \toprule
    \end{tabular}
    \scriptsize{vanilla $=$ no prosody modeling, P $=$ Phoneme-level feature, W $=$ Word-level feature, R $=$ Rule-based prosody labels (i.e. F0 and energy), N $=$ Neural-based prosody label (i.e. VQ-VAE codeword), H=Hierarchical model, rand. $=$ prosody label randomly sampled from the uniform prior of VQ-VAE.}
    \vspace{-1.7mm}
    \caption{Objective prosody scores for TTS models with different prosody-modeling frameworks.}
    \vspace{-3.2mm}
    \label{tab:objective}
\end{table}

\vspace{-1.7mm}
\subsection{Objective evaluation}
\label{ssec:objective_evaluation}
\vspace{-1.2mm}
It is difficult to evaluate the quality and naturalness of synthesized audio samples with objective metrics.
However, there are several metrics used in early works to evaluate pairwise prosody similarity \cite{Skerry-Ryan2018, Chu2009}:
\begin{itemize}
    \vspace{-0.7mm}
    \item Gross Pitch Error (GPE): measuring the pitch similarity between two utterances.
    \vspace{-0.7mm}
    \item Voice Decision Error (VDE): measuring the difference of voiced/unvoiced decision between two utterances.
    \vspace{-0.7mm}
    \item F0 Frame Error (FFE): combining GPE and VDE.
\end{itemize}
For these metrics, We use the DIO algorithm for the F0 estimation and the frame-wise voicing decision of both ground-truth and synthesized utterances.
To match the length of two utterances, Dynamic Time Warping \cite{Kominek2008} is used to align the mel-spectrograms, and the resulted alignment is applied to the F0 values and voicing decisions of both utterances.

In addition to the metrics above, we also use the MAE of F0 and energy, which is more compatible with our training objective, to evaluate the synthesized utterances' prosody.
Let $f_t$, $f'_t$ be the estimated F0, $v_t$, $v'_t$ be the binary voiced/unvoiced decision, $e_t$ and $e'_t$ be the L2 norm of the $t$-th STFT frame (after alignment) of the ground-truth and the synthesized audio sample, and $T$ be the number of frames after alignment. We define two metrics \textit{F-MAE} and
\vspace{-0.7mm}
\textit{E-MAE} as below:
\vspace{-0.7mm}
\begin{align*}
    \textit{F-MAE} &= \frac{\sum_{t=1}^T |f_t-f'_t| \mathbbm{1}[v_t] \mathbbm{1}[v'_t]}{\sum_{t=1}^T \mathbbm{1}[v_t] \mathbbm{1}[v'_t]},\\ \textit{E-MAE} &= \frac{\sum_{t=1}^T |e_t-e'_t|}{T}
\end{align*}
\vspace{-0.7mm}
To reduce the differences caused by vocoding, we use the utterances reconstructed by MelGAN from ground-truth mel-spectrograms as the ground-truth for all the objective metrics.

The result is shown in Table \ref{tab:objective}.
It can be seen that for the non-hierarchical models, word-level features achieve better performance than phoneme-level features (W+* v.s. P+*). The models trained with rule-based prosody labels are slightly better than those trained with neural-based features in most cases (*+R v.s. *+N).
We hypothesize it is because the neural-based labels are too complicated to be accurately predicted.
However, the scores of the utterances synthesized with predicted neural-based labels are still much better than randomly sampled prosody labels (P+N, rand., W+N, rand.).

The scores of the four hierarchical models are generally better than their non-hierarchical counterparts.
However, it is not clear which hierarchical model is the best, or which type of prosody label is the most effective one at which level.
One should note that the metrics used in Table \ref{tab:objective} only measure the similarity between the prosody of the synthesized utterances and the ground truth, not the quality of the audio.
The issue will be addressed in the next subsection.

\vspace{-1.7mm}
\subsection{Subjective evaluation}
\label{ssec:subjective_evaluation}
\vspace{-1.2mm}
We conducted three different subjective assessments, including Mean Opinion Score (MOS), Comparison MOS (CMOS) \cite{Loizou2011} and the AXY test \cite{Skerry-Ryan2018}, for audio quality and prosody naturalness evaluation.
100 sentences randomly sampled from the testing set are used for all the assessments.
In each assessment, every utterance (or utterance pair, for CMOS and AXY) received five ratings from crowdsourced human raters.

\vspace{-1.7mm}
\subsubsection{MOS score}
\label{sssec:mos_score}
\begin{figure}[tb]
\centering
\centerline{\includegraphics[width=8.5cm]{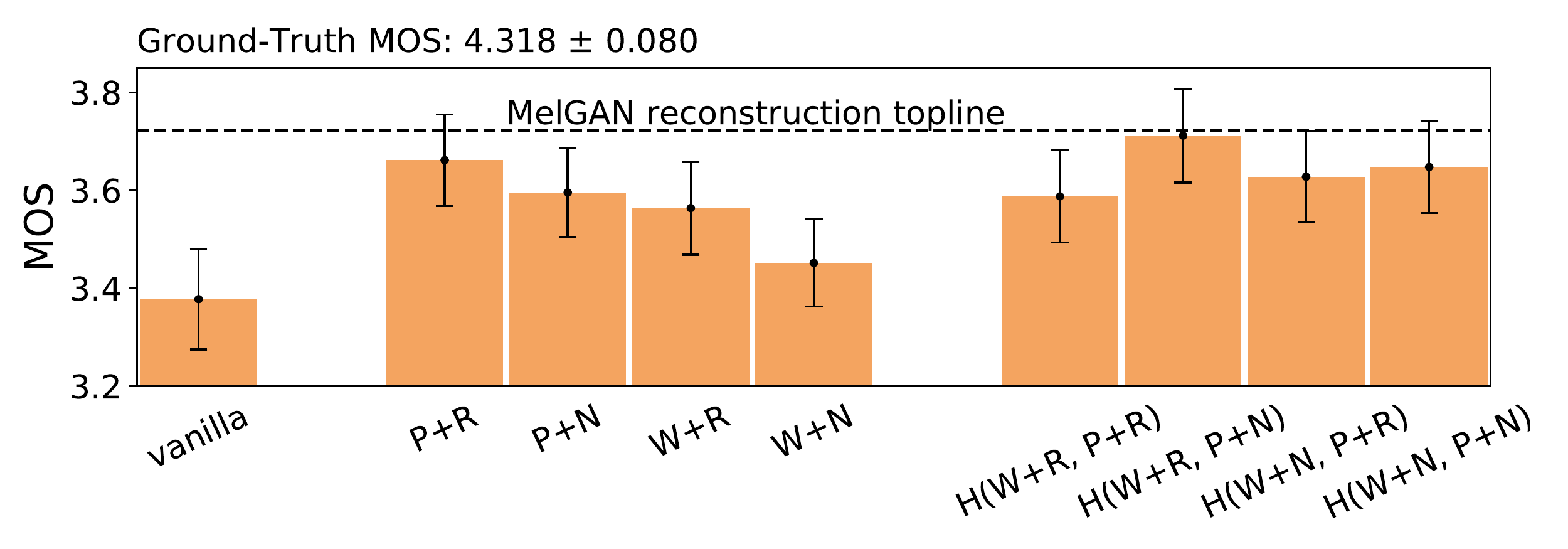}}
\vspace{-4mm}
\caption{5-point scale MOS naturalness evaluation of the TTS models. The error bars indicate the 95 \% confidence intervals.}
\vspace{-2.2mm}
\label{fig:mos}
\end{figure}

\vspace{-1.2mm}
MOS score is used to evaluate the quality of audio samples synthesized by the proposed hierarchical models and other models.
The results are listed in Table \ref{fig:mos}.
All TTS models with prosody modeling achieved better performance than the vanilla baseline.
Looking into the non-hierarchical models, we find that despite the accurate prosody prediction observed in Table \ref{tab:objective}, word-level prosody modeling is still inferior to phoneme-level prosody modeling in terms of the MOS scores.
We carefully inspect the audio samples and find that there are more artifacts and noises in audio samples synthesized with the word-level models. 
The information provided by the coarse-grained prosody labels is not specific enough, hindering the model from generating clear speech.
It can also be seen that the non-hierarchical models with rule-based prosody labels are better than those with neural-based prosody labels.

The MOS scores of the hierarchical models and the phoneme-level models are very close, which implies that the granularity of prosody modeling is important for perceptual audio quality.
The audio samples of the best hierarchical model H(W+R, P+N) are even comparable to the samples reconstructed from the ground-truth mel-spectrograms, reflecting the effectiveness of prosody modeling.

\vspace{-1.7mm}
\subsubsection{Pairwise subjective scores}
\label{sssec:pairwise_subjective_scores}
\vspace{-1.2mm}
We further conduct CMOS and AXY test to compare the hierarchical models and the non-hierarchical models pairwisely.
In the CMOS test, the human raters are asked to select one audio sample with better quality and naturalness out of two.
In the AXY test, the raters are presented two synthesized audio samples and a ground-truth, and they are asked to select one sample with prosody more similar to the ground-truth, regardless of the audio quality and any other factors.
Before the AXY test, four audio samples are used to instruct the raters to ignore audio quality and focus on prosody. 
The hierarchical model with word-level rule-based labels and phoneme-level neural-based labels is used for all the pairwise comparison tests since it outperformed any other hierarchical models in the MOS test.
The results are listed in Table \ref{tab:axy_cmos}.


\begin{table}[tb]
    \centering
    \begin{tabular}{ccccccc}
        \toprule
        && CMOS & $p$-value && AXY & $p$-value\\
        \toprule
        P+R && $.000$ & $.500$ && $.114$ & $.027$\\
        \hline
        P+N && $.190$ & $9.9 \times 10^{-4}$ && $.160$ & $.006$\\
        \hline
        W+R && $.088$ & $.049$ && $.070$ & $.108$\\
        \hline
        W+N && $.302$ & $4.6 \times 10^{-7}$ && $.048$ & $.224$\\
        \toprule
    \end{tabular}
    \vspace{-2.7mm}
    \caption{CMOS score and AXY reference similarity score of the proposed hierarchical model (W+R, P+N) against non-hierarchical models. Both scores are in a -3 $\sim$ 3 scale, with 0 begin neutral and the larger the better. The $t$-test $p$-values are also reported.}
    \vspace{-4.2mm}
    \label{tab:axy_cmos}
\end{table}

The CMOS scores are consistent with the MOS scores, which shows that the hierarchical models are slightly better than the phoneme-level models in terms of naturalness, while the difference is not significant.
The AXY test reflects that the prosody of audio samples synthesized with the proposed hierarchical model is better than the non-hierarchical models, especially the phoneme-level models, which is also observed in Table \ref{tab:objective}.
The prosody prediction of the word-level models is better but the audio quality is worse than the phoneme-level models, also consistent with previous experiments.

\vspace{-1.7mm}
\section{Conclusion}
\label{sec:conclusion}
\vspace{-1.2mm}
The result shows that word-level prosody modeling achieves more accurate prosody prediction than phoneme-level ones, but subjective tests show that the perceptual quality degrades due to coarse granularity.
The proposed hierarchical prosody modeling framework combines the advantages of both word- and phoneme-level, thus capable of generating high-quality audio samples with accurate prosody.


\vspace{-1.7mm}
\section{Acknowledgement}
\label{sec:acknowledgement}
\vspace{-1.2mm}
We thank to National Center for 
High-performance Computing (NCHC) of Taiwan for providing computational and storage resources.

\pagebreak
\bibliographystyle{IEEEbib}
\bibliography{strings,refs}

\end{document}